\documentclass{article}
\usepackage{spconf, amsmath, subfigure, graphicx, amsfonts, amssymb}
\usepackage{multirow, multicol, diagbox, makecell, arydshln, booktabs}
\usepackage{enumitem}
\usepackage{epstopdf}

\usepackage{enumitem}
\usepackage{mathtools}
\usepackage{subfigure}
\usepackage{hyperref}
\usepackage{url}
 \usepackage{balance}

\usepackage{color}

\setlist{nosep, leftmargin=14pt}

\title{Structure-Preserving Graph Kernel for Brain Network Classification}

\name{Jun Yu$^{1}$, Zhaoming Kong$^{1}$, Aditya Kendre$^{2}$, Hao Peng$^{3}$, Carl Yang$^{4}$, Lichao Sun$^{1}$, Alex Leow$^{5}$, Lifang He$^{1}$}
\address{$^{1}$Department of Computer Science and Engineering, Lehigh University, PA, USA\\
$^{2}$Cumberland Valley High School, PA, USA\\
$^{3}$School of Cyber Science and Technology, Beihang University, Beijing, China\\
$^{4}$Department of Computer Science, Emory University, GA, USA\\
$^{5}$Department of Psychiatry, University of Illinois at Chicago, IL, USA}

\newcommand{\ourmethod}[0]{$\text{SSGK}_{\text{w/o sparse}}$}
\newcommand{\ours}[0]{$\text{SSGK}$}

\newtheorem{theo}{\textsc{Theorem}}

\begin{document}
%
\maketitle
\begin{abstract}
Brain network analysis is of great importance in clinical diagnosis and treatments. In this paper, we present a novel graph-based kernel learning approach for brain network classification. Specifically, we demonstrate how to exploit the natural graph structure of brain networks to encode prior knowledge in the kernel using the tensor product operator. For each brain network, we first proposed to apply sparse matrix factorization with a symmetric constraint to extract tensor product based approximation. We then used them to derive a structure-persevering symmetric graph kernel to be fed into the support vector machine (SVM). Quantitative evaluations on challenging EEG-based emotion recognition tasks with respect to different frequency bands demonstrate the superior performance of our proposed method, compared with the state-of-the-art traditional and deep learning methods. Together, results show that relevant EEG signals are primarily encoded in the alpha and theta bands during the emotion regulation task, which is consistent with previous findings.
\end{abstract}



%
\vspace{-5pt}
\begin{figure*}[htpb]
\centerline{\includegraphics[width=0.6 \linewidth]{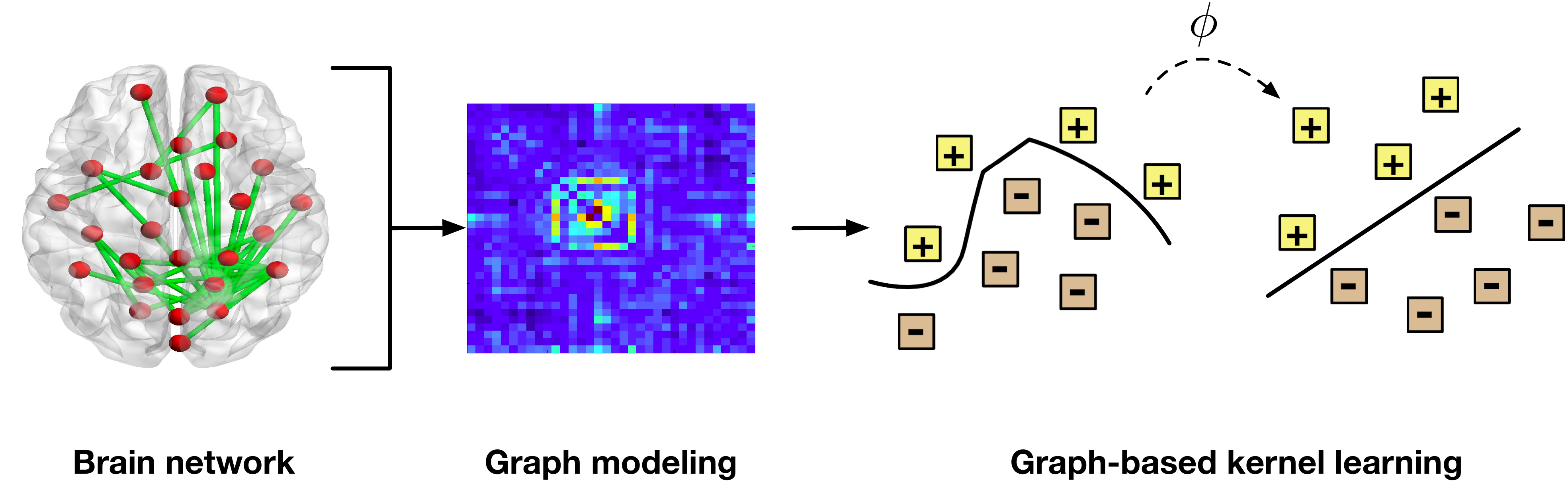}}
\caption{The framework of graph-based kernel learning.}
\label{fig:frame}
\vspace{-10pt}
\end{figure*}

\section{Introduction} \label{sec:intro}
Brain network analysis, enriched by the advances of neuroimaging technologies such as electroencephalography (EEG) and diffusion tensor imaging (DTI), has been an appealing research topic in recent years in neuroscience~\cite{fornito2016fundamentals}. The study originates from modeling the human brain connectome as a graph -- a mathematical construct mapping the connectivity of anatomically distinct brain regions (i.e., nodes) and inter-regional pathways (i.e., edges). 
By graph based analysis, the information encoded by the connectome can promote critical understanding on how the brain manages cognition, what signals the connections convey and how these signals affect brain regions~\cite{zhang2019new}. It has shown great potential in disease diagnosis, clinical outcome prediction, therapeutic adjustment and collection of biological features~\cite{wang2014human,gao2015identifying,yahata2016small}. With the development of machine learning algorithms on graph-structured data, it is of great importance to apply such approaches to brain network analysis. In particular, it is desirable to develop more accurate predictive methods as a complement to the effort of pathologists in diagnosis process and treatment decision-making.

In the past decades, a variety of machine learning methods have been explored for brain network selection and classification. For example, support vector machine (SVM)~\cite{vapnik2013nature}, graph kernel~\cite{yang2020unified}, frequent graph-based pattern mining (gSpan)~\cite{du2016network}, tensor decomposition~\cite{leonardi2013identifying, he2014dusk}. Deep learning methods such as convolution neural network (CNN)~\cite{wang2017structural} and graph convolutional network~\cite{zhang2018multi}, which are successful on many tasks, are exploited as well. Although great achievements have been made in various research aspects of these methods, some issues still exists. The human connectome has complex and non-linear characteristics, which may not be well captured by linear models. Meanwhile, deep learning methods suffer from the enormous parameter sizes, which is both difficult for training and vulnerable to overfitting. Besides, many methods do not make good use of graph structure. Thus, it is desirable to develop a concise method for brain network analysis.

In this paper, we propose a novel graph-based kernel learning approach for brain network predictive analysis, and apply it to the challenging EEG-connectome emotion regulation task. The contributions of this work are threefold:
\begin{itemize}
\item We derived a structure-preserving symmetric graph kernel (SSGK) in tensor product space for brain network classification. A new matrix factorization scheme was introduced to incorporate the graph structure as well as the symmetric constraint and sparse layouts. 
\item Extensive experiments on multiclass EEG-based emotion regulation task with respect to different frequency bands demonstrate the superior performance of SSGK, compared with the state-of-the-art traditional and deep learning methods. Results also show that relevant EEG signals are primarily encoded in alpha and theta bands during the emotion regulation task, which is consistent with previous studies.
\item SSGK is a general graph-kernel framework for efficiently measuring the similarity of structured data. It has great potentials for a wide range of applications, in conjunction with various kernel-based methods and kernel functions.
\end{itemize}

\section{Materials and Methods} \label{sec_prelim}
In this section, we introduce notations and basic concepts, and then describe our method in detail.

\subsection{Notations and Concepts.}


Following~\cite{kolda2009tensor}, we denote vectors by lowercase boldface letters, e.g., $\mathbf{x}$; and matrices by uppercase boldface, e.g., $\mathbf{X}$. An index is denoted with a lowercase letter, spanning the range from 1 to the uppercase letter of the index, e.g., $i=1, 2, \cdots, I$. We denote a matrix as $\mathbf{A} \in \mathbb{R}^{I \times J}$, and their elements by $a_{i,j}$. We will often use calligraphic letters ($\mathcal{A}$, $\mathcal{B}$, $\mathcal{C}$, $\cdots$) to denote general space. The inner product of two matrices $\mathbf{A}, \mathbf{B} \in \mathbb{R}^{I \times J}$ is defined as $\langle \mathbf {A}, \mathbf {B} \rangle=\sum_{i=1}^{I}\sum_{j=1}^{J} a_{i,j}b_{i,j}$. A rank-one matrix $\mathbf{A}$ equals to the outer product of two vectors: $\mathbf{A}=\mathbf{u} \otimes \mathbf{v},~ \mbox{where} ~ a_{i,j} = u_i v_j$. Note that for rank-one matrices it holds that 
\vspace{-6pt}
\begin{equation}\label{eq4}
\langle \mathbf{a} \otimes \mathbf{b}, \mathbf{u} \otimes \mathbf{v} \rangle =\langle \mathbf{a}, \mathbf{u} \rangle \langle \mathbf{b}, \mathbf{v} \rangle.
\end{equation}
\textbf{Kernel learning.} Support vector machines (SVMs) are one of the most popular kernel-based learning algorithms, which are effective on the data by linear boundaries, and kernel functions are adopted to classify by non-linear boundaries~\cite{vapnik2013nature}. The kernel function encapsulates the hypothesis language, i.e., how to perform data transformation and knowledge encoding. In general, it maps data from the original input feature space to a higher dimensional feature space (known as Hilbert space), and a kernel function corresponds to the inner product in this higher dimensional feature space. The computational attractiveness of kernel methods comes from the fact that quite often a closed form of `feature space inner products' exists~\cite{gartner2003survey}. Instead of mapping the data explicitly, the kernel can be calculated directly. According to \emph{Mercer's theorem}~\cite{vapnik2013nature}, we can verify whether a kernel function is valid by the following Theorem~\cite{berlinet2011reproducing}.
\vspace{-2pt}
\begin{theo} \label{theo1}
A function $\kappa$ defined on $\mathcal{X} \times \mathcal{X}$ is a positive definite kernel of $\mathcal{H}$ if and only if there exists a feature mapping function $\phi(\cdot): \mathcal{X} \mapsto \mathcal{H}$ such that $\kappa (\mathbf{x}, \mathbf{y}) = \langle \phi (\mathbf{x}), \phi (\mathbf{y}) \rangle$, for any $(\mathbf{x}, \mathbf{y}) \in \mathcal{X} \times \mathcal{X}$.
\end{theo}
\vspace{-2pt}

In particular, an important property of positive definite kernels is that they are closed under sum,
multiplication by a scalar and product~\cite{cristianini2000introduction}.

\subsection{Structure-Preserving Symmetric Graph Kernel} \label{sec:method}
The brain networks are biologically expected to be both sparse and highly localized in space. Such unique characterizations put specific topological constraints onto machine learning models we can use effectively. We propose a new matrix factorization scheme to incorporate the graph structure as well as the symmetric constraint and sparse layouts, which allows one to interpret brain network as a bilinear tensor product approximation. We then use this approximation to define a structure-preserving symmetric graph kernel function (SSGK) for the SVM classifier. The framework of the proposed method is illustrated in Figure \ref{fig:frame}, and we present the key steps of our methods in detail as below.



\textbf{Feature extraction.} The graph provides a natural representation for connectome data, but there is no guarantee that such representation will be good for kernel learning. Since learning will only be successful if the regularities that underlie the data can be discerned by the kernel. From the characteristics of connectome objects, we know that the essential information in the connectome is embedded in the structure of the graph. Thus, one important aspect of kernel learning for such complex objects is to represent them by sets of key structural features which are easier to manipulate. In previous work~\cite{he2014dusk}, it was found that matrix factorization is particularly effective for extracting this structure. It can take the correlation in the graph matrix into account and represent it directly into a sum of rank one matrices (bilinear bases), yielding a more compact representation of connectome data. Motivated by these observations, we use matrix factorization for feature extraction. In particular, given a graph matrix $\mathbf{X} \in \mathbb{R}^{I \times I}$, we investigate the following optimization problem:
\vspace{-5pt}
\begin{equation}
\underset{\mathbf{a}_r}{\min} \| \mathbf{X} - \sum_{r=1}^{R} \mathbf{a}_r \otimes \mathbf{a}_r \|_F^2 + \lambda \sum_{r=1}^{R}\|\mathbf{a}_r \|_1,
\label{eq:mf}
\end{equation}
where $R$ is the rank of the matrix $\mathbf{X}$ defined as the smallest number of rank-one matrices in an exact matrix factorization, $\|\cdot\|_F$ is the Frobenius norm of a matrix, and $\|\cdot\|_1$ is the $\ell_1$ norm, also known as the lasso regularization for a sparse solution. Equation (\ref{eq:mf}) can be solved by the tensorlab toolbox\footnote{https://www.tensorlab.net/}.

\textbf{Graph structure mapping.} Although matrix factorization operation decomposes the graph matrix, the graph structure can still be preserved and retrieved based on the factorized results. We show how the above feature extraction results can be exploited to induce a structure-preserving graph kernel. Suppose we are given the matrix factorization of $\mathbf{X}, \mathbf{Y} \in \mathbb{R}^{I \times I}$ by $\mathbf{X}= \sum_{r=1}^{R} \mathbf{x}_{r} \otimes \mathbf{x}_{r}$ and $\mathbf{Y}= \sum_{r=1}^{R} \mathbf{y}_{r} \otimes \mathbf{y}_{r}$, respectively. We assume the graph observations are mapped into the Hilbert space $\mathcal{H}$ by
\vspace{-5pt}
\begin{equation}\label{eq9}
\phi: \mathbf{X} \rightarrow \phi \left(\mathbf{X}\right) \in \mathbb{R}^{\mathcal{H} \times \mathcal{H}}.
\end{equation}
Importantly, the mapping result $\phi(\mathbf{X})$ is still a symmetric matrix, but its dimension is higher than $\mathbf{X}$, even infinite depending on the feature mapping function $\phi(\cdot)$.
 
\begin{figure}[t]
	\centering
	\includegraphics[width=0.8 \linewidth]{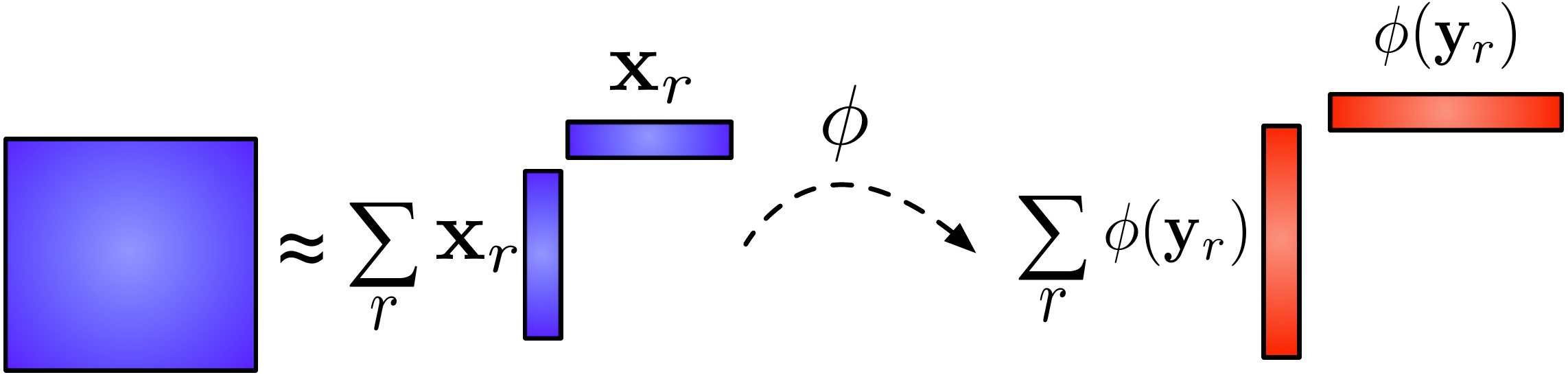}
    \caption{Schemic diagram of the feature extraction and graph structure mapping}
	\label{fig:method}
	\vspace{-10pt}
\end{figure}

Based on the definition of the kernel function, we know that the feature space is a high-dimensional space generated from the original space, equipped with the same operations. Thus, we can factorize graph data directly in the feature space in the same way as in the original space. This is formally equivalent to performing the following mapping:
\begin{equation}\label{eq22}
\phi: \sum_{r=1}^{R} \mathbf{x}_{r} \otimes \mathbf{x}_{r} \rightarrow \sum_{r=1}^{R} \phi(\mathbf{x}_{r}) \otimes \phi(\mathbf{x}_{r}).
\end{equation}
In this sense, it corresponds to mapping graphs into high-dimensional graphs that retain the original structure. More precisely, it can be regarded as mapping the original graph matrix to matrix feature space and then conducting the matrix factorization in the feature space, as illustrated in Fig.~\ref{fig:method}.

After mapping the matrix factorization into the outer product feature space, the kernel can be defined directly with the inner product in that feature space. Thus, based on equation (\ref{eq4}), we can derive our SSGK model:
\vspace{-5pt}
\begin{equation}
\label{eq23}
\begin{split}
    \kappa(\mathbf{X},\mathbf{Y}) & = \kappa (\sum_{r=1}^{R} \mathbf{x}_r \otimes \mathbf{x}_r, \sum_{r=1}^{R} \mathbf{y}_r \otimes \mathbf{y}_r) \\
    & = \big \langle \sum_{r=1}^{R} \phi(\mathbf{x}_{r}) \otimes \phi(\mathbf{x}_{r}), \sum_{r=1}^{R} \phi(\mathbf{y}_{r}) \otimes \phi(\mathbf{y}_{r}) \big \rangle \\
    & = \sum_{p=1}^{R} \sum_{q=1}^{R} \kappa (\mathbf{x}_p,\mathbf{y}_q)\kappa (\mathbf{x}_p,\mathbf{y}_q).
\end{split}
\end{equation}


Based on Theorem \ref{theo1}, it is not difficult to see that this kernel is `valid' as it is described as an inner product of two matrices $\sum_{r=1}^{R} \phi(\mathbf{x}_{r}) \otimes \phi(\mathbf{x}_{r})$ and $\sum_{r=1}^{R} \phi(\mathbf{y}_{r}) \otimes \phi(\mathbf{y}_{r})$. From the derivation process, we know that such a kernel can take into account the flexibility of graph structure. In general, SSGK is an extension of the conventional kernels in the vector space to matrix space, and each vector kernel can be used in this framework for EEG-connectome analysis in conjunction with kernel machines. Our positive result can be viewed as saying that designing a good graph kernel function is much like designing a good graph structure in the feature space.

\section{Experiments}\label{sec:exp}

\textbf{Participants and data acquisition.} The dataset used in this paper were collected from 22 healthy participants at the University of Illinois at Chicago (UIC) and from 11 healthy participants at the University of Michigan (UMich). Each participant underwent an Emotion Regulation Task (ERT). During the ERT session, participants were instructed to look at pictures displayed on the screen. Emotionally neutral pictures (e.g., landscape, everyday objects) and negative pictures (e.g., car crash, nature disasters) would appear on the screen for seven seconds in random orders. One second after the picture on display, a corresponding auditory guide would instruct the participant to {\em neutral}: viewing the neutral pictures; to {\em maintain}: viewing the negative pictures as they normally would; or to {\em reappraise}: viewing the negative pictures while attempting to reduce their emotion response by re-interpreting the meaning of pictures. All subjects were recorded using the Biosemi system equipped with an elastic cap with 34 scalp channels. The acquisition connectivity matrix is $34 \times 34$ with 130 time points and 50 frequencies ranging from 1Hz to 50Hz in increments of 1Hz. A detailed description about data acquisition and preprocessing is available in~\cite{xing2016eeg}. 

\begin{figure}[htbp]
	\label{fig:heatmap}
	\centering
	\includegraphics[width=0.78 \linewidth]{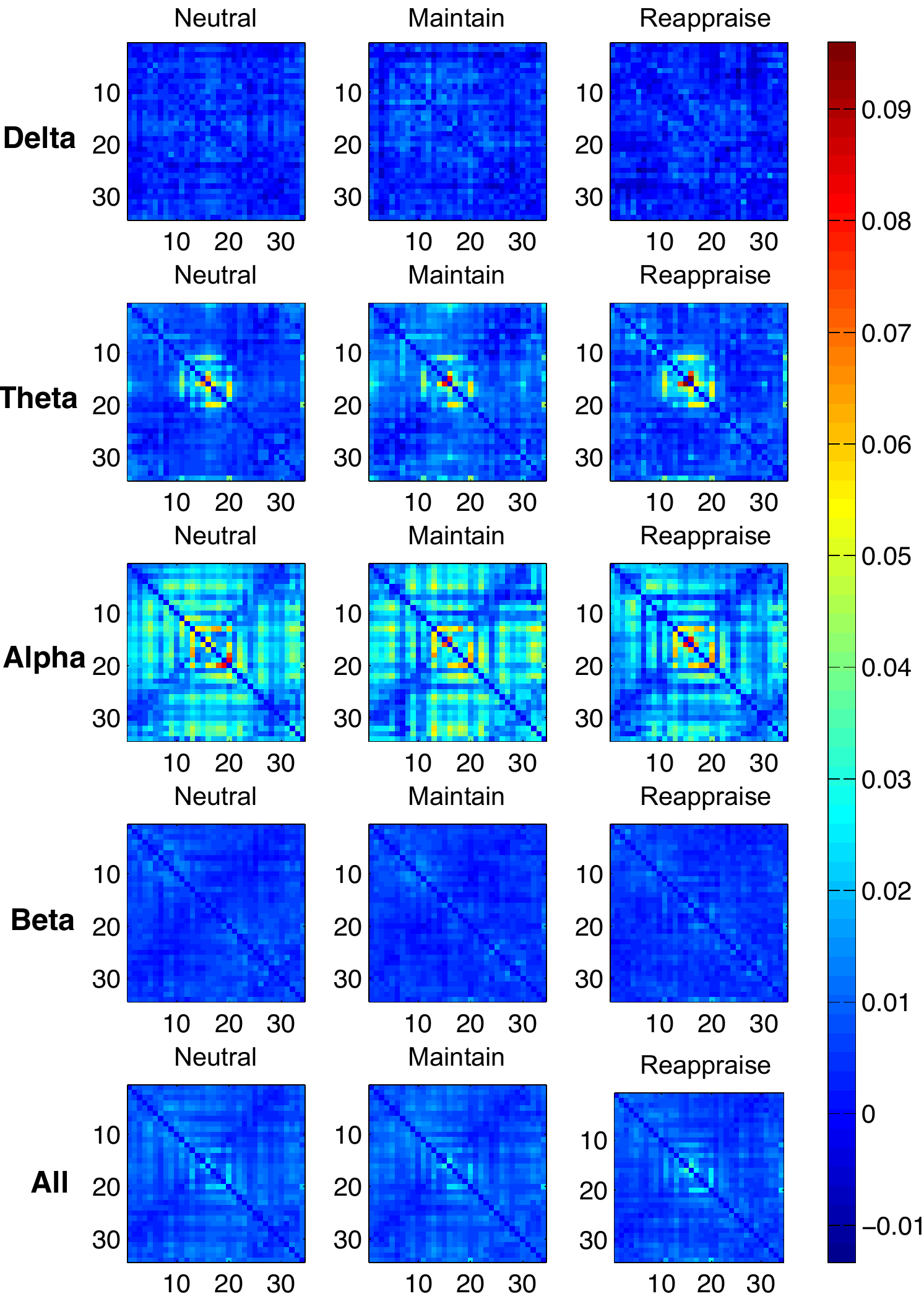}
	\vspace{-10pt}
    \caption{Average EEG-connectome during neutral, maintain and reappraise in the five different frequency bands.}
    \vspace{-15pt}
\end{figure}

\textbf{Tasks.} We study multi-class EEG-connectome emotion regulation tasks and analyze the effect of different frequency bands of EEG signals. In emotion regulation, studies have shown that relevant EEG information is primarily encoded in the low frequency bands~\cite{balconi2015hemodynamic}. Thus, we analyze the EEG-connectome data in 5 frequency bands: Delta (1--3 Hz), Theta (4--7 Hz), Alpha (8--12 Hz), Beta (13--30 Hz), for relative power, as well as the total power of the EEG (1--30 Hz)~\cite{lutz2009mental}. The average EEG-connectome during neutral, maintain and reappraise in the five different frequency bands are shown in Figure~\ref{fig:heatmap}, where the $x$-- and $y$--axes represent the vertex id, and the color of the cell represents the strength of the connectivity between vertices $x$ and $y$. We can see that the connectivity in the alpha band is generally stronger than other frequency bands, and theta band is the second one.

\textbf{Compared methods.} We evaluate eight algorithms in Table \ref{tab:exp} on the five tasks above, each of which represents a different strategy: the edge based feature extraction (Edge), where edge values are directly used as features by flatting connectivity matrices of EEG-connectome into vectors; the local clustering coefficients (CC)~\cite{rubinov2010complex}, which measures a network’s local segregation; the characteristic path length (CPL)~\cite{watts1998collective} that quantifies the global information integration; the graph-based substructure pattern mining (gSpan)~\cite{yan2002gspan} as a discriminative subgraph selection approach; the dual structure-preserving kernel (DuSK)~\cite{he2014dusk}, which takes multidimensional tensors as input. We use second- (i.e., averaged over time and frequency), third- (i.e., averaged over time), and fourth-order (i.e., all data with dimension $34 \times 34 \times 130 \times x$, where $x$ corresponds to a number of the frequency level) version of this scheme, denoted as DuSK-2D, DuSK-3D and DuSK-3D, respectively; the convolutional neural network (CNN) with 2D convolutions for averaged 2D brain network data and 3D convolutions for averaged 3D brain network data~\cite{gupta2013natural}; the graph convolutional network (GCN) for averaged 2D brain network data~\cite{zhang2018multi}, where the average of all brain networks is used as the graph structure (i.e., adjacency matrix) for information propagation; the proposed method and its variant without sparse-constraint ({\ours} and {\ourmethod}).

\textbf{Experimental settings.} In our experiments, we use the subjects collected from UIC as the training set (66 samples), and UMich as the testing set (33 samples). Following~\cite{he2014dusk}, we choose SVM with Gaussian RBF kernel as the classifier for all methods. Classification accuracy is used as the evaluation metric. All compared methods select the optimal trade-off parameter of SVM and kernel width from $ \{2^{-8}, 2^{-7}, \cdots, 2^{8}\}$. Other parameters for gSpan and DuSK are set according to~\cite{yan2002gspan} and~\cite{he2014dusk}, respectively. For our {\ours} and {\ourmethod} methods, the parameter $R$ and $\lambda$ was selected from the value set of $R = \{1, 2, \cdots, 12\}$ and  $\lambda =  \{2^{-2}, 2^{-1}, \cdots, 2^{8}\}$ using grid search.

\textbf{Results.} Detailed results of compared methods are listed in Table \ref{tab:exp}. From Table \ref{tab:exp}, it can be seen that the proposed SSGK-based methods outperform all compared methods by 10\%- 20\% on almost all five different frequency bands. The superiority of the proposed methods demonstrate the effectiveness of utilizing the structure information within the graph representation during encoding. Specifically, among all five bands, SSGK produces the best performance on Alpha band and second best performance on Theta band, which is consistent with previous findings~\cite{xing2016eeg} and can also be observed in our visualization in Figure 3. Furthermore, by comparing SSGK and {\ourmethod}, it is noticed that the proposed SSGK approach with sparse regularization consistently outperforms the same approach without sparse regularization, and the advantage of sparsity characterization indicates the importance of modeling the redundant information of observed frequency bands.

\begin{table}[t]
\scriptsize
\vspace{-10pt}
\caption{The classification accuracy in percentage (\%) by competing methods and the proposed method for five tasks. The best results for each task are highlighted in boldfont.} 
\label{tab:exp}
\centering
\scalebox{0.93}{
\begin{tabular}{c|c|ccccc}
\toprule
&  & \multicolumn{5}{c}{Frequency Band} \\
\cmidrule{3-7}
Category & ~~~Method~ & ~~Delta~   & ~~Theta~  & ~~Alpha~  & ~~Beta~  & ~~All~~~ \\
\midrule 
\multirow{8}*{Traditional} &  Edge        &42.42    & 54.55   & 51.52   & 51.52   & 45.45 \\
\cline{3-7}
& CC         &54.55    & 54.55   & 42.42   & 51.52   & 42.42\\
\cline{3-7}
& CPL        &48.48    & 42.42   & 45.45   & 48.48   & 39.39\\
\cline{3-7}
& gSpan      & 39.39   & 51.52   & 39.39   & 54.55   & 48.48 \\
\cline{3-7}
& DuSK--2D  &51.52    & 63.64   & 51.51   & 51.52   & 54.55  \\
\cline{3-7}
& DuSK--3D  &57.58    & 57.58   & 57.58   & 54.55   & 48.48 \\
\cline{3-7}
& DuSK--4D  &54.55    & 54.55   & 51.52   & 54.55   & 57.58 \\
\midrule
\multirow{4}*{Deep Learning} & CNN--2D  &51.11   &43.71  &43.07  &42.54   &41.48 \\
\cline{3-7} 
& CNN--3D  &46.67  &45.93  &41.48  &57.04  &44.44   \\
\cline{3-7}
& GCN  & 41.31   & 48.08   & 41.01  &  40.61 & 37.37 \\
\midrule
\multirow{2}*{Ours} & {\ourmethod} &57.58 & 66.67 & 63.64  & 54.55  & 57.58 \\
& {\ours} & \textbf{63.64}  & \textbf{69.70} & \underline{\textbf{72.73}}  & \textbf{60.61} & \textbf{57.58} \\
\bottomrule
\end{tabular}
}
\vspace{-10pt}
\end{table}

\vspace{-5pt}
\section{Conclusion}
\vspace{-5pt}
This paper proposes a graph-based kernel learning approach called Structure-preserving Symmetric Graph Kernel (SSGK) for brain network classification task. The proposed method mainly follows two consecutive steps: first, a sparse-inducing symmetric matrix factorization strategy is applied to extract structural features from the natural symmetric graph representations of the brain network data, then the extracted structural features are directly used to define the SSGK function and further fed into the support vector machine for the classification.
Experimental results on challenging EEG-based emotion recognition task demonstrates the effectiveness of the proposed method for encoding prior knowledge in the kernel using structural information of brain networks. 

\section{Acknowledgements}
This work was funded in part by ONR Grant N00014-18-1-2009 and Lehigh's accelerator grant S00010293 to Lifang He.

\vfill
\pagebreak

\section{Compliance with Ethical Standards}
This work is related to human studies. The EEG dataset used in this study are owned by a third-party organization, where informed consent was obtained for all subjects. All studies are conducted according to the Good Clinical Practice guidelines and U.S. 21 CFR Part 50 (Protection of Human Subjects), and under the approval of Institutional Review Boards.

\bibliographystyle{IEEEbib}

\end{document}